\documentclass{kluwer}    
\usepackage{epsfig} 

\newdisplay{guess}{Conjecture}

\def\gray{$\gamma$-ray\ }
\def\grays{$\gamma$-rays\ }
\def\Berat{$^{10}$Be$/\,^9$Be}
\def\Alrat{$^{26}$Al$/\,^{27}$Al}

\def\Ferat{sub-Fe/Fe}

\def\Dunits{$\times10^{28}$ cm$^2$ s$^{-1}$}

\def\vunits{km s$^{-1}$}

\newcommand{\jref}[3]{{\it #1\/} {\bf #2}, #3}
\def\ocircle{~\put(0,2){\tiny $\bigcirc$}\hspace*{1em}}
\def\obox{~\put(0,4){\fbox{}}\hspace*{1em} }
\def\fcircle{{\Large $\bullet$} } 
\def\odiamond{{ $\diamond$} }
\def\otriangle{{\footnotesize $\triangle$} }
\def\fha{80mm}
\def\fwa{118mm}
\def\fwb{58mm} 

\begin{document}                                                                                   
\begin{article}
\begin{opening}

\title{Diffuse Galactic $\gamma$-rays: \\
Constraining Cosmic-Ray Origin and Propagation}

\author{Igor V.~\surname{Moskalenko}\thanks{also
Institute for Nuclear Physics, Moscow State University, Moscow, 
Russia}
and Andrew W.~\surname{Strong}}
\institute{
Max-Planck-Institut f\"ur extraterrestrische Physik, D--85740 Garching, Germany}

\runningauthor{Igor V.~Moskalenko and Andrew W.~Strong}
\runningtitle{Constraining the Cosmic-Ray Origin and Propagation}
\date{\today}

\begin{abstract}
We have developed a model which aims to reproduce observational data of
many kinds related to cosmic-ray (CR) origin and propagation: direct
measurements of nuclei, antiprotons, electrons and positrons,
$\gamma$-rays, and synchrotron radiation.  Our main results include
evaluation of diffusion/convection and reacceleration models, estimates
of the halo size, calculations of the interstellar positron and
antiproton spectra, evaluation of alternative hypotheses of nucleon and
electron interstellar spectra, and computation of the Galactic diffuse
\gray emission.  
Recently our CR propagation code has been generalized
to include fragmentation networks of arbitrary complexity.  The code
can now provide an alternative to leaky-box calculations for full
isotopic abundance calculations and has the advantage of including the
spatial dimension which is essential for radioactive nuclei.
Preliminary predictions  for \Ferat, \Berat\ and \Alrat\ 
are presented in anticipation of new experimental isotopic
data.  We show that combining information from classical CR
studies with \gray and other data leads to tighter constraints on
CR origin and propagation.
\end{abstract}

\end{opening}

\section{Introduction and the Modelling Approach}  

A numerical method for the calculation of Galactic CR
propagation in 3D has been developed.  
Our program\footnote{Our model (``GALPROP'') including software and
datasets is available at
http://www.gamma.mpe--garching.mpg.de/$\sim$aws/aws.html }
(\citeauthor{SM98,MS98,MSR98,SMR99}) performs CR propagation
calculations for nuclei, antiprotons, electrons and positrons and
computes \gray and synchrotron emission in the same framework.  The 3D
spatial approach with a realistic distribution of interstellar gas
distinguishes it from leaky-box calculations.   The basic spatial
propagation mechanisms are diffusion and convection, while in momentum
space energy loss and diffusive reacceleration are treated.
Fragmentation and energy losses are computed using realistic
distributions for the interstellar gas and radiation fields.  The basic
procedure is first to obtain a set of propagation parameters which
reproduce the CR B/C and \Berat\ ratios; the same propagation
conditions are then applied to all the CR species. Gamma-ray and
synchrotron emission are then evaluated with the same model.
We aim for a ``standard
model'' which can be improved with new astrophysical input and
additional observational constraints.

GALPROP solves the Galactic CR propagation equation numerically on a
grid in 3D with cylindrical symmetry, and the basic coordinates are
$(R,z,p)$, where $R$ is Galactocentric radius, $z$ is the distance from
the Galactic plane, and $p$ is the total particle momentum.  In the
models the propagation region is bounded by $R=R_h$, $z=\pm z_h$ beyond
which free escape is assumed.  For a given $z_h$ the diffusion
coefficient as a function of momentum is determined by B/C for the case
of no reacceleration; if reacceleration is assumed then the
reacceleration strength (related to the Alfv\'en speed, $v_A$) is
constrained by the energy-dependence of B/C.  We include diffusive
reacceleration since some stochastic reacceleration is inevitable, and
it provides a natural mechanism to reproduce the energy dependence of
the B/C ratio without an {\it ad hoc} form for the diffusion
coefficient (e.g., \opencite{SeoPtuskin94}).  The distribution of CR
sources is chosen to reproduce (after propagation) the CR distribution
determined by analysis of EGRET \gray data \cite{StrongMattox96}.  The
bremsstrahlung and inverse Compton (IC) \grays are computed
self-consistently from the gas and radiation fields used for the
propagation.

Tighter constraints on the parameters of CR propagation and source
abundances can be obtained from consideration of all CR isotopes
simultaneously.  We have now been
able to generalize the scheme to include the CR reaction networks of
arbitrary complexity.  The scheme can hence potentially compete with
leaky-box calculations for computation of isotopic abundances, while
retaining  the spatial component essential for radioactive nuclei,
electrons and $\gamma$-rays.  Even for stable nuclei it has the advantage
of a physically-based propagation scheme with a spatial distribution of
sources rather than an {\it ad hoc} path length distribution.  It also
facilitates tests of reacceleration.

The new extended approach for nuclei handles the reaction network
explicitly (\citeauthor{SM99b}), as follows:
(i) propagate primary species from an assumed set of source abundances,
(ii) compute the resulting spallation source function for all species,
(iii) propagate all species using the primary and spallation sources.
Steps (ii) and (iii) are iterated until converged.
After the second iteration the result is already accurate for the pure
secondary component, after the third iteration it is accurate for
tertiaries, and so on.  Hence in practice only a few such iterations
are necessary for a complete solution of the network.  The method is
more time-consuming than the simpler approach we used before, where
multiple progenitors and tertiary etc. reactions were handled by
weighting the cross-sections, since many more species are included and
because of the several iterations required.

\section{Probes of Interstellar Propagation}

Measurements of nucleons in CR provide a basis for probing interstellar
propagation and determining the halo size.  The ratio of stable
secondary to primary nuclei is often used to constrain such parameters
as the diffusion coefficient, reacceleration strength and/or convection
velocity, while long-lived radioactive species can be a probe of the
extent of the propagation volume.

To constrain the parameters of propagation we use the B/C ratio which
is measured over a wide energy range. 
In diffusion/convection models with a diffusion coefficient which is a
simple power-law in momentum a good fit is {\it not} possible; the
basic effect of convection is to reduce the variation of B/C with
energy, and although this improves the fit at low energies the
characteristic peaked shape of the measured B/C cannot be reproduced.
Although modulation makes the comparison with the low energy
data somewhat uncertain, the fit is unsatisfactory.  
We concluded therefore that convection does not seem to 
work in detail, but requires an artificial break in the diffusion
coefficient.  If instead we consider diffusive reacceleration, the peak
is produced rather naturally (as many people have pointed out) so that
we rather prefer this solution.

Propagation parameters\footnote{  The diffusion coefficient, adjusted
to fit B/C, differs slighly from that used in the original work for the
same $z_h$ (\citeauthor{SM98}), reflecting the more detailed treatment
and updated cross sections.  } for a model with reacceleration are:
halo size $z_h = 4$ kpc, $v_A = 20$ \vunits, diffusion coefficient
$D=D_0\beta{(p/p_0)}^{1/3}$, where $D_0 = 6.75$\Dunits, and $p_0 = 3$
GeV.

\begin{figure}[t]
\psfig{file=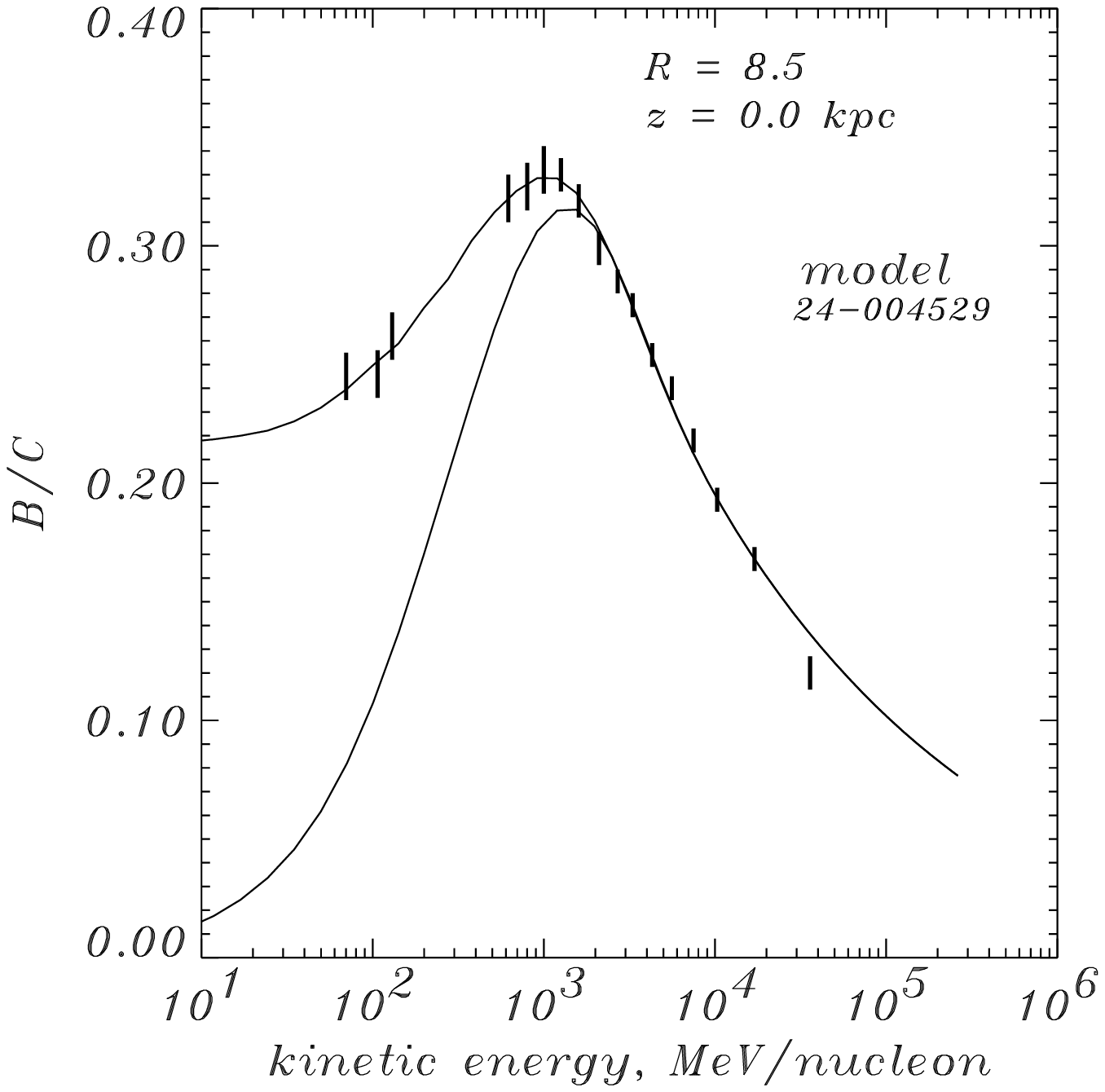,width=\fwb} \hfill 
\psfig{file=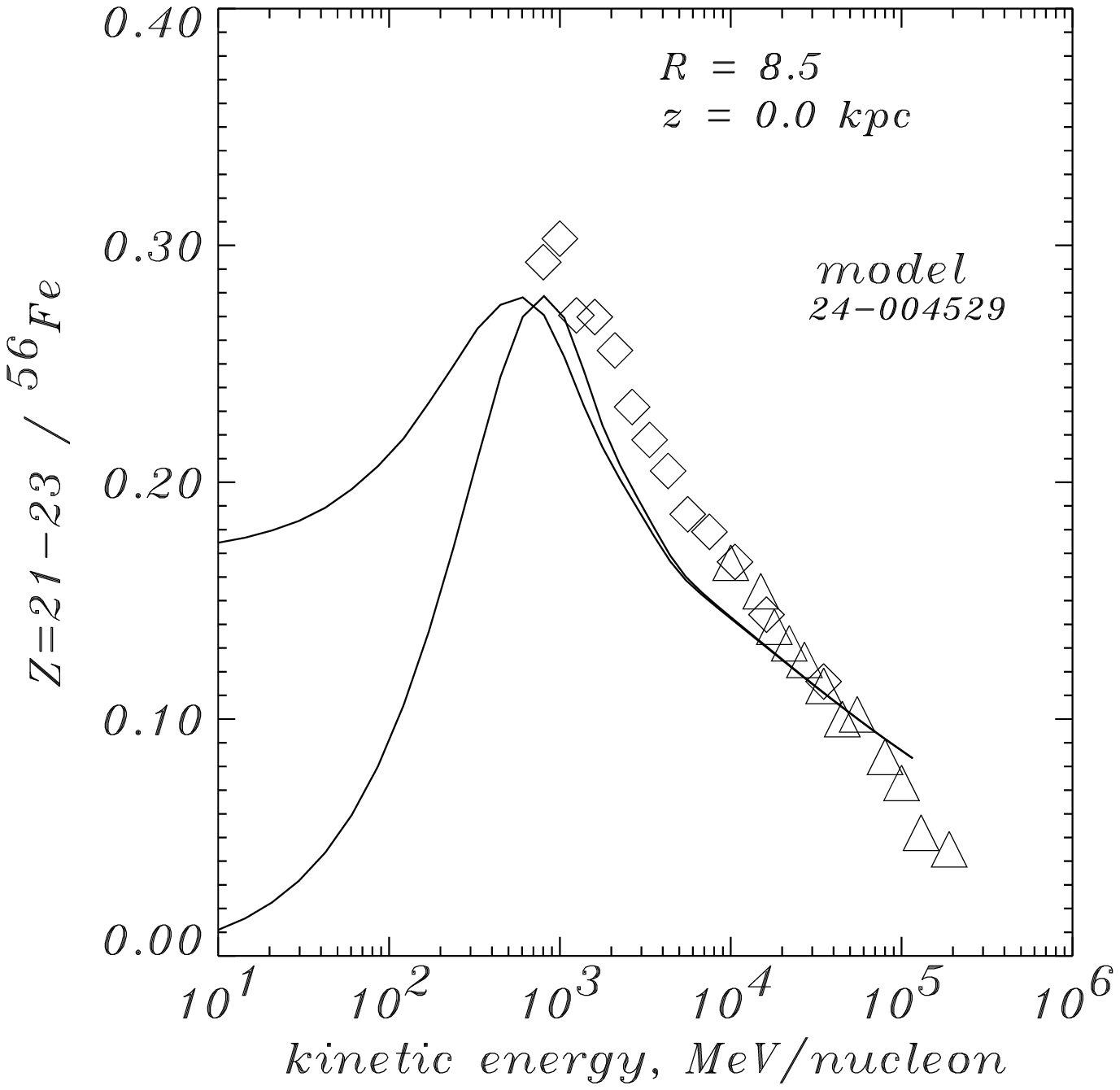,width=\fwb}
\caption[]{ {\it Left}: B/C interstellar and modulated to 500 MV for
diffusive reacceleration model with $z_h$ = 4 kpc.  Data compilation:
\inlinecite{Webber96}.
{\it Right}: The same for \Ferat.  Data:  \odiamond --
\inlinecite{Engelmann90}, \otriangle -- \inlinecite{Binns88}. }
\label{fig1}
\end{figure}

\begin{figure}[t]
\psfig{file=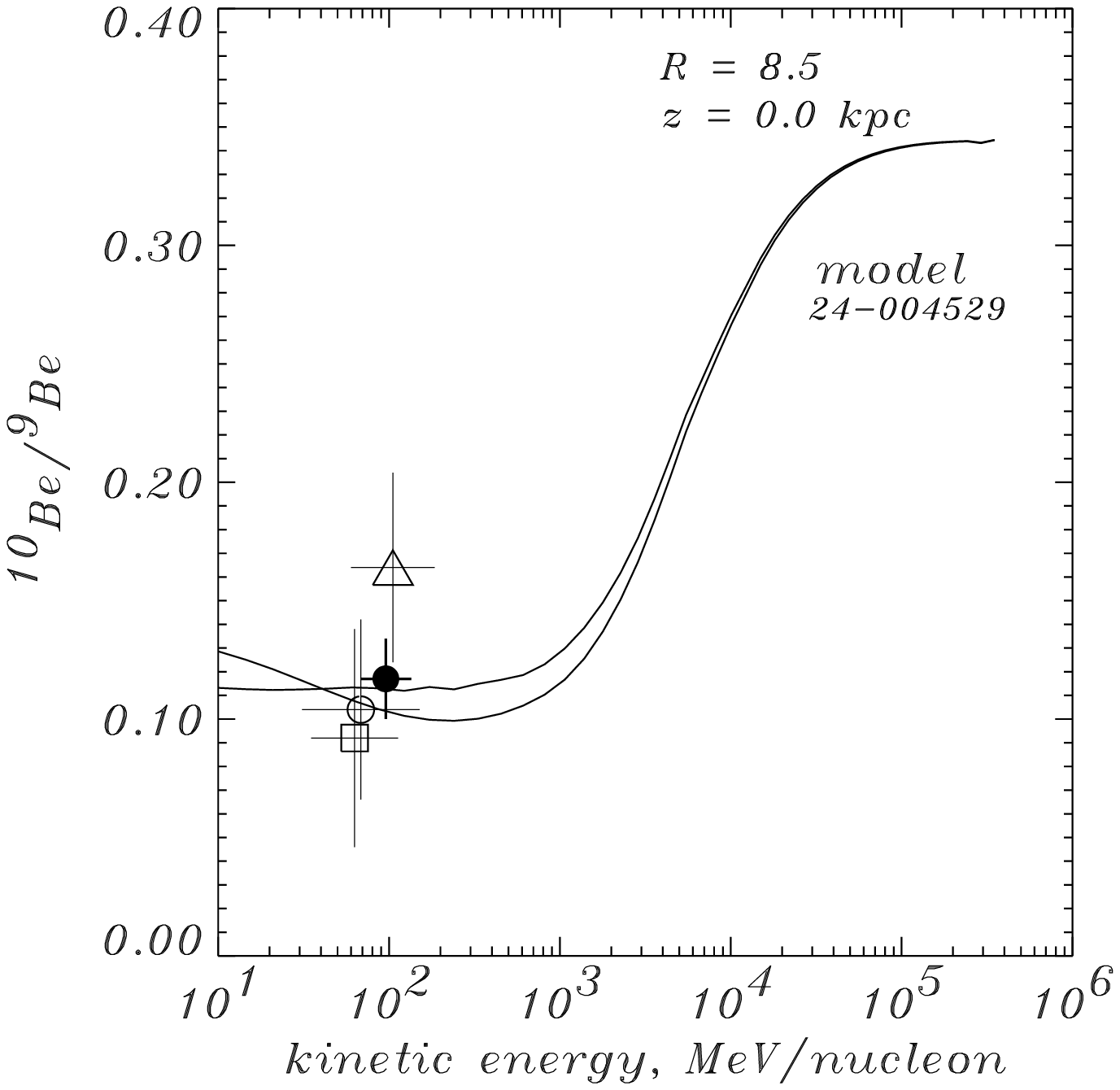,width=\fwb} \hfill 
\psfig{file=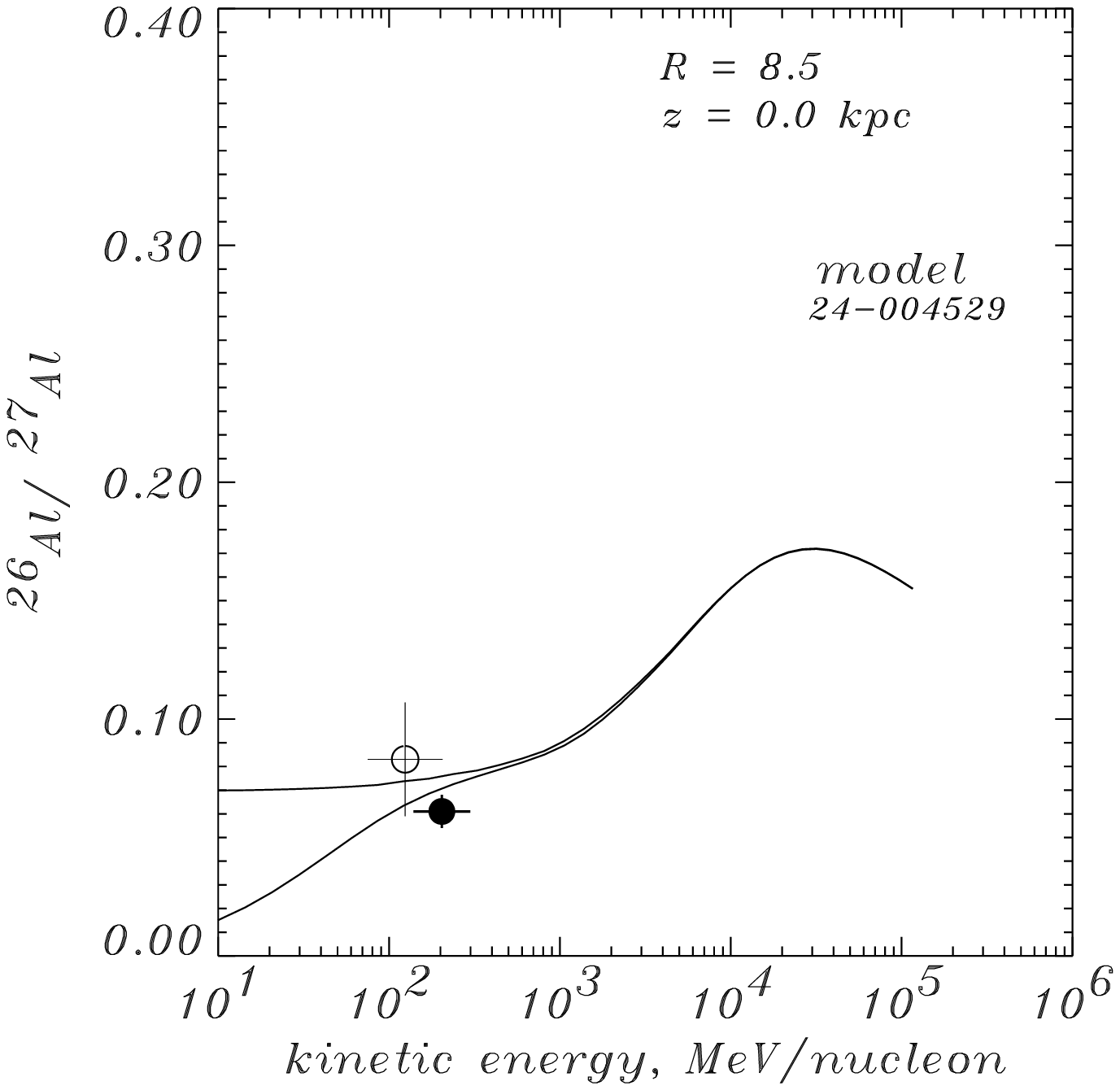,width=\fwb}
\caption[]{ Model interstellar and modulated (500 MV) ratios.  
{\it Left}: \Berat. Data from \inlinecite{Lukasiak94a} (\obox --
Voyager--1,2, \ocircle -- IMP--7/8, \otriangle -- ISEE--3) and 
\inlinecite{Connell98} (\fcircle -- Ulysses).
{\it Right}: \Alrat. Data: \ocircle -- \inlinecite{Lukasiak94b},
\fcircle -- \inlinecite{SimpsonConnell98}.  Note that the data points
shown are at the measured (not interstellar) energies. }
\label{fig2}
\end{figure}

We have applied the method to a network of 87 nuclei from protons to
Ni, including explicitly all stable species and radioactive species
with half-life more than $10^5$ years.  Isotopic cross sections are
based on measured values where available. Otherwise we use the
\inlinecite{Webber90} cross-section code.  Source abundances are from
\inlinecite{DuVernoisThayer96}, with solar isotopic ratios within a
given element.

Fig.~\ref{fig1} shows B/C and \Ferat.  This model with reacceleration
reproduces the sub-Fe data reasonably well considering that the model
was adjusted only to fit B/C.  The deviation from the observed \Ferat\
shape is however noticeable, as discussed by \inlinecite{Webber97}.

Fig.~\ref{fig2} shows examples of radioactive species.  Modulation is
for nominal values of the modulation parameter but this has little
effect on the comparison due to the small energy dependence of these
ratios in the 100--1000 MeV/nucleon range.  In \citeauthor{SM98} we
obtained a range 4 kpc $<z_h<$ 12 kpc for the halo height; in the present
improved model the \Berat\ and \Alrat\ predictions are consistent with
this.

Recently, \inlinecite{WebberSoutoul98} and
\inlinecite{PtuskinSoutoul98} have obtained $z_h= 2-4$ kpc and
$4.9_{-2}^{+4}$ kpc, respectively, and our results are consistent 
with these.

\begin{figure}[t]
\psfig{file=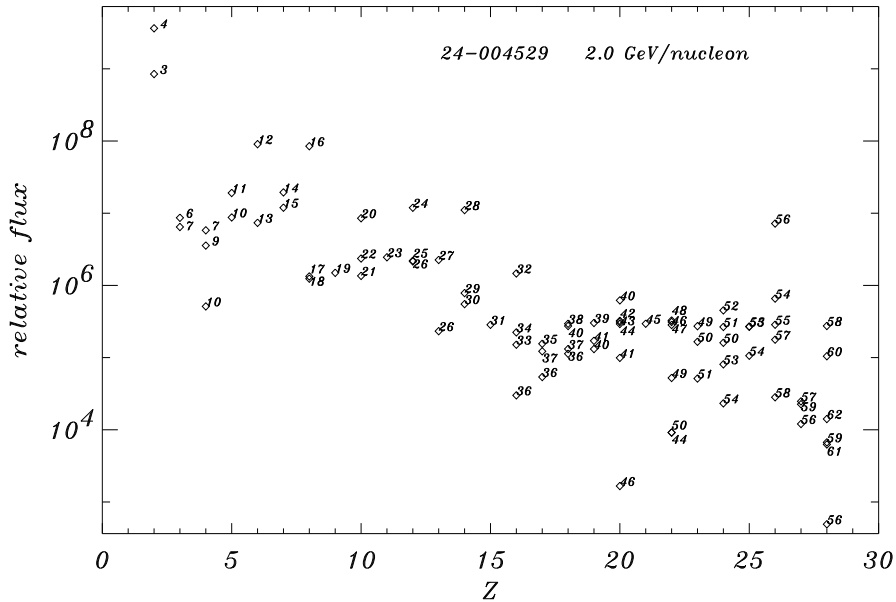,width=\fwa,height=\fha}
\caption[]{ Model isotopic fluxes at 2 GeV/nucleon (\citeauthor{SM99b}) as
function of $Z$; individual $A$ values are marked. }
\label{fig3}
\end{figure}

Fig.~\ref{fig3} shows computed fluxes of
all the included isotopes.   This result is
illustrative of the method but not to be taken as predictions for
evaluation purposes. 

\section{Probes of the Interstellar Nucleon Spectrum: \\
Diffuse Continuum $\gamma$-rays, Antiprotons and Positrons}

Secondary $e^+$'s and $\bar{p}$'s in Galactic CR and some part of the
diffuse Galactic $\gamma$-rays are produced in collisions of CR
particles with interstellar matter\footnote{ Contributions from
possible nearby source(s) of positrons (e.g., \opencite{Aharonian95})
and WIMP annihilation (e.g., \opencite{Bottino98};
\opencite{BaltzEdsjo98}; \citeauthor{MS99}) are also discussed.  }.
Because they are secondary, they reflect the {\it large-scale} nucleon
spectrum independent of local irregularities in the primaries and thus
provide an essential check on propagation models and also on the
interpretation of diffuse \gray emission
(\citeauthor{MS98,MSR98,SMR99}).  These are an important diagnostic for
models of CR propagation and provide information complementary to that
provided by secondary nuclei.  However, unlike secondary nuclei,
$\bar{p}$'s and $e^+$'s reflect primarily the propagation history of
the protons, the main CR component.

The most direct probe of the interstellar proton spectrum is perhaps
provided by diffuse $\gamma$-rays, but an essential and {\it a priori}
unknown part of the emission is produced by CR electrons via IC
scattering.  The latter depends on many details of propagation in the
Galaxy as well as distributions of the magnetic and radiation fields.
Moreover, because of large electron energy losses the average electron
spectrum spectrum in the Galaxy can differ substantially from what is
measured locally.

Recent results from both COMPTEL and EGRET indicate that IC scattering
is a more important contributor to the diffuse emission than previously
believed.  The puzzling excess in the EGRET data $>1$ GeV relative to
that expected for $\pi^0$-decay has been suggested to orginate in IC
scattering from a hard interstellar electron spectrum (e.g.,
\opencite{PohlEsposito98}; \citeauthor{SM99a}), or from a harder
average nucleon spectrum in interstellar space than that observed
directly (e.g., \opencite{Mori97}).  Our combined approach allows us to
test these hypotheses (\citeauthor{SMR99}).

\begin{figure}[t]
\psfig{file=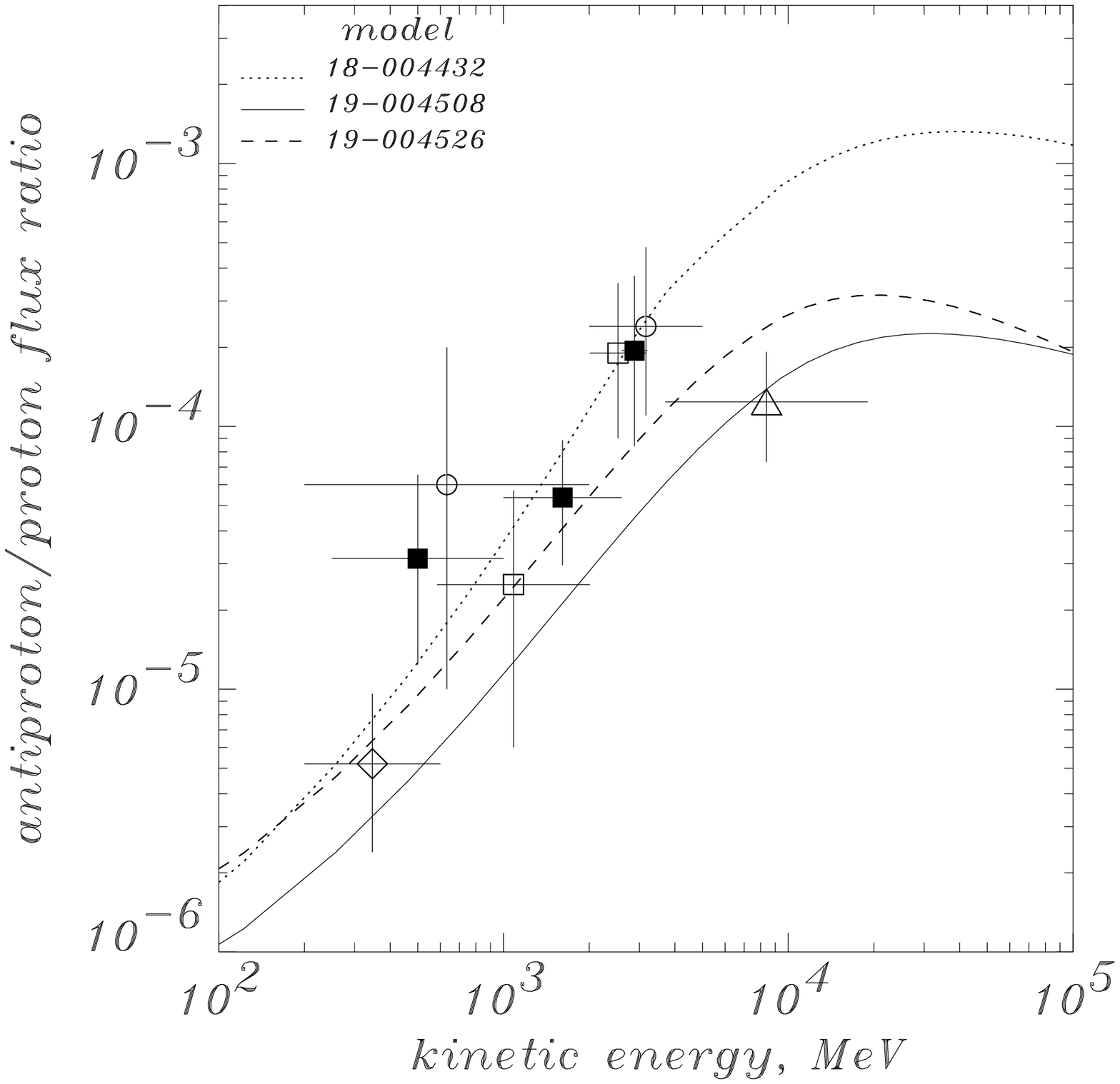,width=\fwb} \hfill 
\psfig{file=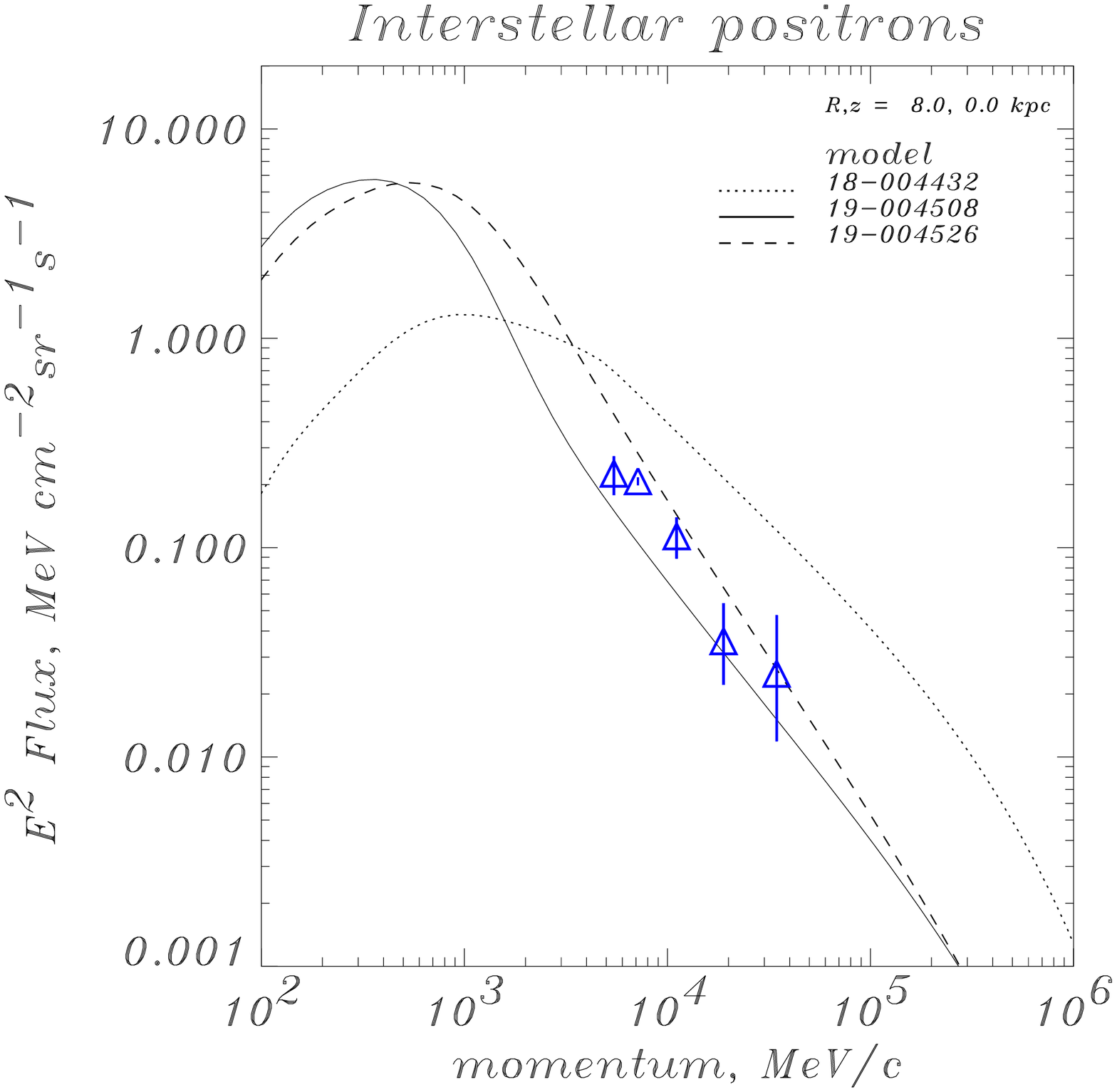,width=\fwb}
\caption[]{
{\it Left:} Interstellar $\bar{p}/p$ ratio for different ambient proton
spectra (\citeauthor{MSR98,SMR99}) compared with data.  Solid line:
``conventional'', dotted line: ``hard nucleons'', dashed line: ``best
model''.  Data (direct measurements): see references in
\citeauthor{MSR98}.
{\it Right:} Interstellar spectra of $e^+$'s for different ambient
proton spectra (\citeauthor{SMR99}) compared with data.  Lines are
coded as on the left.  Data (direct measurements):
\inlinecite{Barwick98}. }
\label{fig4}
\end{figure}

Our calculations show that the suggestion of a hard nucleon spectrum
provides better agreement with EGRET measurements, but conflicts with
data by producing too much $\bar{p}$ and $e^+$
(Fig.~\ref{fig4}).  The hard electron spectrum hypothesis looks more
plausible but still the agreement with \gray data is not too good.

\begin{figure}[t]
\psfig{file=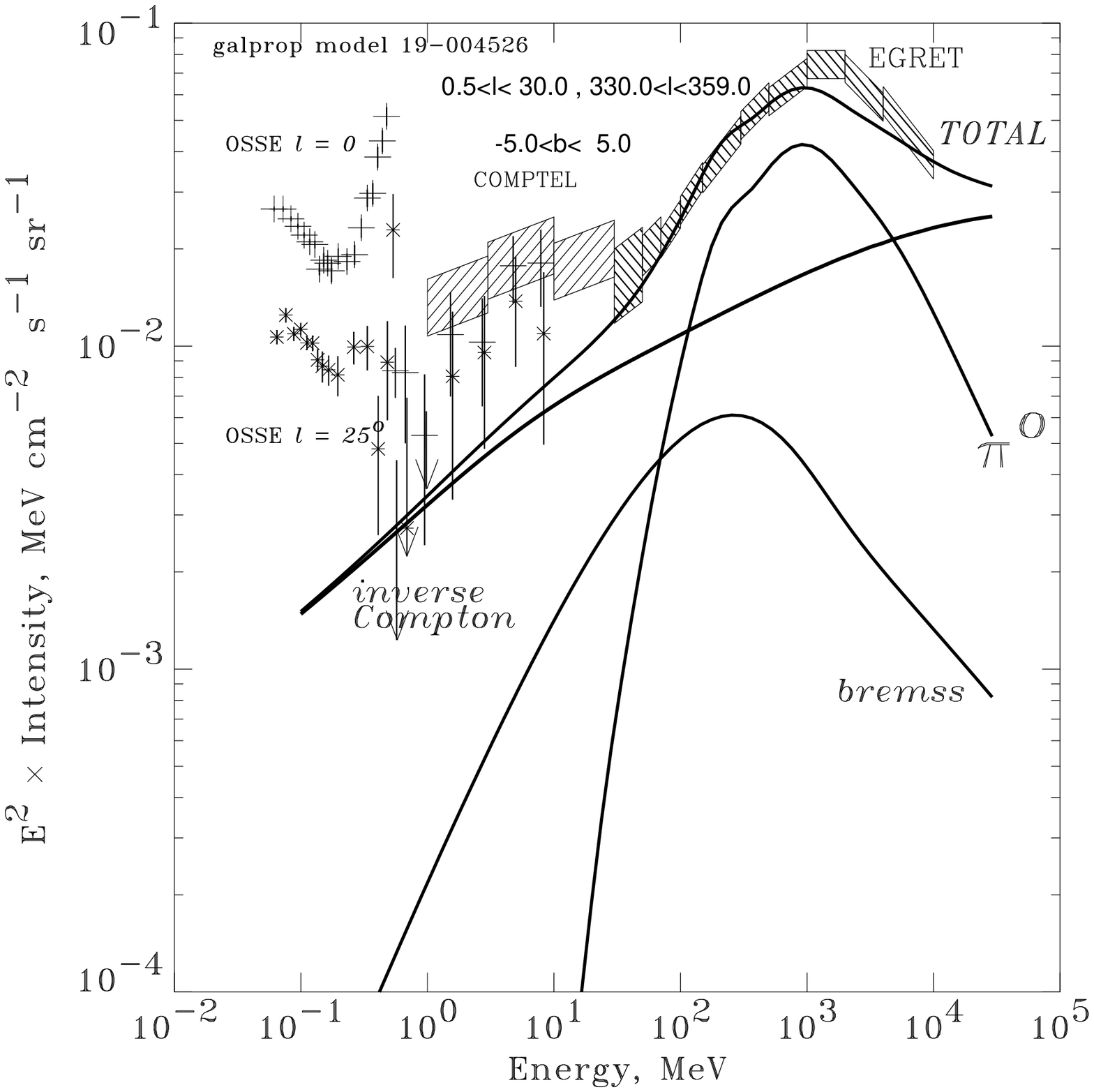,width=\fwb} \hfill 
\psfig{file=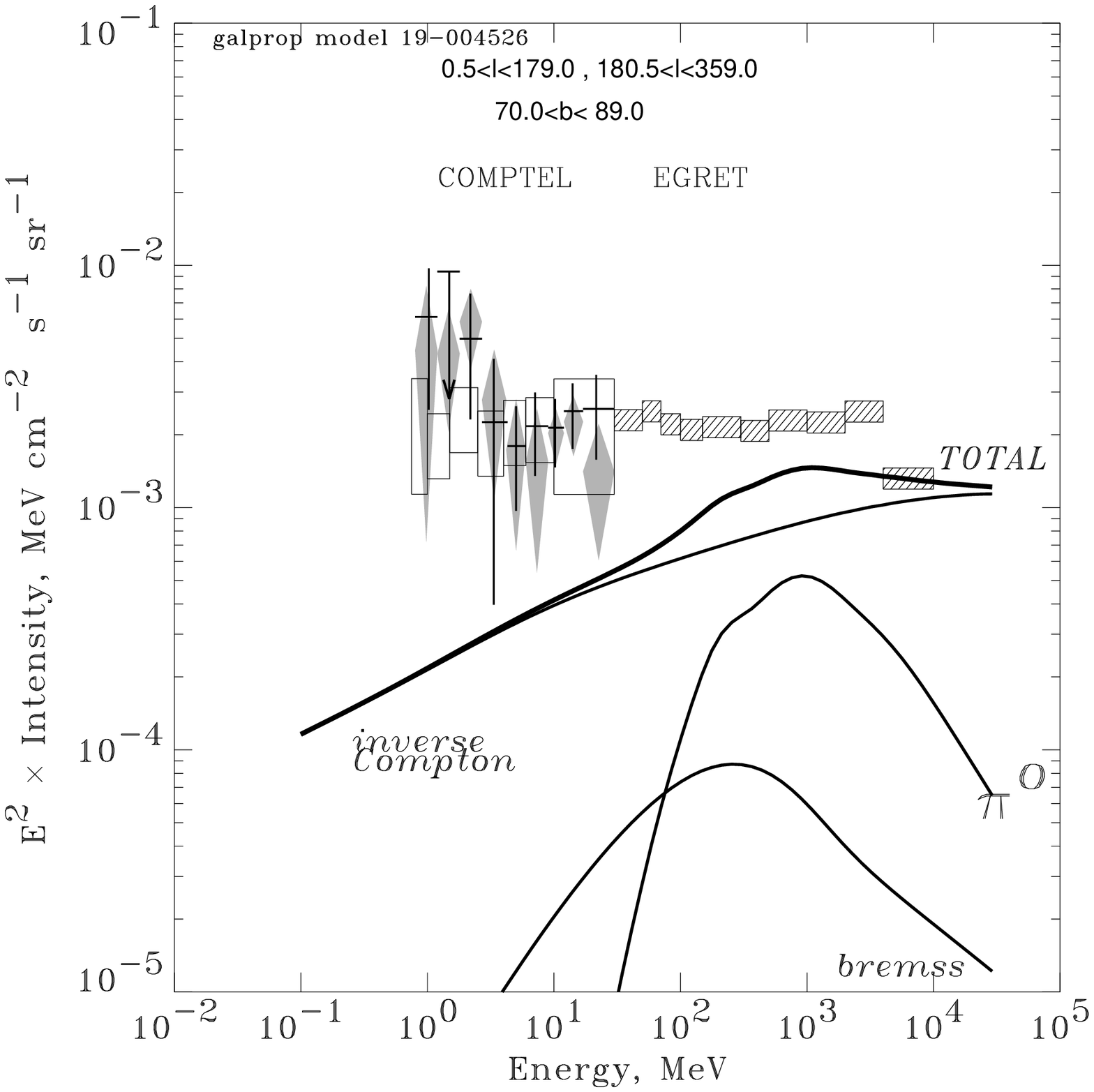,width=\fwb}
\caption[]{
{\it Left:} Gamma-ray energy spectrum of the inner Galaxy ($300^\circ
\le l\le 30^\circ$, $|b|\le 5^\circ$) compared with our best model
calculations ($z_h = 4$ kpc).  Curves show the contribution of IC,
bremsstrahlung, and $\pi^0$-decay, and the total.  Data: EGRET
\cite{StrongMattox96}, COMPTEL \cite{Strong99}, OSSE ($l=0, 25^\circ$:
\opencite{Kinzer99}).
{\it Right:} Energy spectrum of \grays from high Galactic latitudes
($|b|\ge 70^\circ$, all longitudes). Data: see references in
\citeauthor{SMR99}. }
\label{fig5}
\end{figure}

Our best model so far includes the electron injection spectral index
--1.8, which after propagation with reacceleration provides consistency
with radio synchrotron data (a crucial constraint).   Following
\inlinecite{PohlEsposito98}, for this model we do {\it not} require
consistency with the locally measured electron spectrum above 10 GeV
since the rapid energy losses cause a clumpy distribution so that this
is not necessarily representative of the interstellar average.  For
this case, the interstellar electron spectrum deviates strongly from
that locally measured, and also the nucleon spectrum at low energies is
slightly modified to obtain an improved fit to the \gray data.  Because
of the increased IC contribution at high energies, the predicted \gray
spectrum can reproduce the overall intensity from 30 MeV -- 10 GeV
(Fig.~\ref{fig5}) (\citeauthor{SMR99}).

Our calculations of the antiproton/proton ratio and spectra of
secondary $e^+$'s for this model (with reacceleration) are shown in
Fig.~\ref{fig4}.  The predictions are larger than the conventional
model, in which the electron and nucleon spectra are adjusted to agree
with local measurements, but are still consistent with the $\bar{p}$
and $e^+$ measurements.
The analysis of the latitude and longitude \gray distributions 
shows that such a model with large IC component can
indeed reproduce the data (\citeauthor{SMR99}).

None of our models fits the \gray spectrum below $\sim$30 MeV as
measured by the Compton Gamma-Ray Observatory (Fig.~\ref{fig5}).  This
is the consequence of our requirement to be consistent with synchrotron
data.  In order to fit the low-energy part as diffuse emission, without
violating synchrotron constraints (\citeauthor{SMR99}), requires a
rapid upturn in the CR electron spectrum below 200 MeV.  However, in
view of the energetics problems \cite{Skibo97}, a population of
unresolved sources seems a more probable origin for the emission and
would be the natural extension of the low energy plane emission seen by
OSSE \cite{Kinzer99} and GINGA \cite{Yamasaki97}.


\end{article}

\begin{thebibliography}{}

\bibitem[\protect\citeauthoryear{Aharonian et al.}{1995}]{Aharonian95}
   Aharonian, F.A., Atoyan, A.M. and V\"olk, H.J.: 1995, \jref{A\&A}{294}{L41}

\bibitem[\protect\citeauthoryear{Baltz and Edsj\"o}{1998}]{BaltzEdsjo98}
   Baltz, E.A. and Edsj\"o, J.: 1998, \jref{Phys.\ Rev.\ D}{59}{023511}

\bibitem[\protect\citeauthoryear{Barwick et al.}{1998}]{Barwick98}
   Barwick, S.W., et al.: 1998, \jref{ApJ}{498}{779}

\bibitem[\protect\citeauthoryear{Binns et al.}{1988}]{Binns88}
   Binns, W.R., et al.: 1988, \jref{ApJ}{324}{1106} 

\bibitem[\protect\citeauthoryear{Bottino et al.}{1998}]{Bottino98}
   Bottino, A., et al.: 1998, \jref{Phys.\ Rev.\ D}{58}{123503}

\bibitem[\protect\citeauthoryear{Connell}{1998}]{Connell98}
   Connell, J.J.: 1998, \jref{ApJ}{501}{L59}

\bibitem[\protect\citeauthoryear{DuVernois and Thayer}{1996}]{DuVernoisThayer96}
   DuVernois, M.A. and Thayer, M.A.: 1996, \jref{ApJ}{465}{982}

\bibitem[\protect\citeauthoryear{Engelmann et al.}{1990}]{Engelmann90}
   Engelmann, J.J., et al. 1990, \jref{A\&A}{233}{96}

\bibitem[\protect\citeauthoryear{Kinzer et al.}{1999}]{Kinzer99}
   Kinzer, R.L., Purcell, W.R. and Kurfess, J.D.: 1999, \jref{ApJ}{515}{215}

\bibitem[\protect\citeauthoryear{Lukasiak et al.}{1994a}]{Lukasiak94a}
   Lukasiak, A., et al.: 1994a, \jref{ApJ}{423}{426}

\bibitem[\protect\citeauthoryear{Lukasiak et al.}{1994b}]{Lukasiak94b}
   Lukasiak, A., et al.: 1994b, \jref{ApJ}{430}{L69}

\bibitem[\protect\citeauthoryear{Mori}{1997}]{Mori97}
   Mori, M.: 1997, \jref{ApJ}{478}{225}

\bibitem[\protect\citeauthoryear{MS98}{1998}]{MS98}
   Moskalenko, I.V. and Strong, A.W.: 1998, \jref{ApJ}{493}{694} (MS98)

\bibitem[\protect\citeauthoryear{MSR98}{1998}]{MSR98}
   Moskalenko, I.V., Strong, A.W. and Reimer O.: 1998, \jref{A\&A}{338}{L75}
   (MSR98)

\bibitem[\protect\citeauthoryear{MS99}{1999}]{MS99}
   Moskalenko, I.V. and Strong, A.W.: 1999, {\it Phys.\ Rev.\ D}, in press
   (MS99)

\bibitem[\protect\citeauthoryear{Pohl and Esposito}{1998}]{PohlEsposito98}
   Pohl, M. and Esposito, J.A.: 1998, \jref{ApJ}{507}{327}

\bibitem[\protect\citeauthoryear{Ptuskin and Soutoul}{1998}]{PtuskinSoutoul98}
   Ptuskin, V.S. and Soutoul, A.: 1998, \jref{A\&A}{337}{859}

\bibitem[\protect\citeauthoryear{Seo and Ptuskin}{1994}]{SeoPtuskin94}
   Seo, E.S. and Ptuskin, V.S.: 1994, \jref{ApJ}{431}{705}

\bibitem[\protect\citeauthoryear{Simpson and Connell}{1998}]{SimpsonConnell98}
   Simpson, J.A. and Connell, J.J.: 1998, \jref{ApJ}{497}{L85}

\bibitem[\protect\citeauthoryear{Skibo et al.}{1997}]{Skibo97}
   Skibo, J.G., et al.: 1997, \jref{ApJ}{483}{L95}

\bibitem[\protect\citeauthoryear{Strong and Mattox}{1996}]{StrongMattox96}
   Strong, A.W. and Mattox, J.R.: 1996, \jref{A\&A}{308}{L21}

\bibitem[\protect\citeauthoryear{SM98}{1998}]{SM98}
   Strong, A.W. and Moskalenko, I.V.: 1998, \jref{ApJ}{509}{212} (SM98)

\bibitem[\protect\citeauthoryear{SM99a}{1999a}]{SM99a}
   Strong, A.W. and Moskalenko, I.V.: 1999a, In Ramaty, R., et al., editors,
   {\it Proc.\ Workshop LiBeB, Cosmic Rays and Gamma-Ray
   Line Astronomy}, \jref{ASP Conf.~Ser.}{171}{162}, Astron.\ Soc.\ Pacific
   (SM99a)

\bibitem[\protect\citeauthoryear{SM99b}{1999b}]{SM99b}
   Strong, A.W. and Moskalenko, I.V.: 1999b, {\it 26th ICRC (Salt Lake City)}, 
   OG 3.2.18 (SM99b)

\bibitem[\protect\citeauthoryear{SMR99}{1999a}]{SMR99}
   Strong, A.W., Moskalenko, I.V. and Reimer O.: 1999, {\it ApJ}, submitted,
   astro-ph/9811296 (SMR99)

\bibitem[\protect\citeauthoryear{Strong et al.}{1999}]{Strong99}
   Strong, A.W., et al.: 1999, {\it Astroph.\ Lett.\ Comm.}, in press

\bibitem[\protect\citeauthoryear{Webber}{1997}]{Webber97}
   Webber, W.R.: 1997, \jref{Spa.\ Sci.\ Rev.}{81}{107}

\bibitem[\protect\citeauthoryear{Webber and Soutoul}{1998}]{WebberSoutoul98}
   Webber, W.R. and Soutoul, A.: 1998, \jref{ApJ}{506}{335}

\bibitem[\protect\citeauthoryear{Webber et al.}{1990}]{Webber90}
   Webber, W.R., Kish, J.C. and Schrier, D.A.: 1990, 
   \jref{Phys.\ Rev.\ C}{41}{566}

\bibitem[\protect\citeauthoryear{Webber et al.}{1996}]{Webber96}
   Webber, W.R., et al.: 1996, \jref{ApJ}{457}{435}

\bibitem[\protect\citeauthoryear{Yamasaki et al.}{1997}]{Yamasaki97}
   Yamasaki, N.Y., et al.: 1997, \jref{ApJ}{481}{821}

\end{thebibliography}
\end{document}